\documentclass[12pt]{article}

\renewcommand{\baselinestretch}{1.5}

\newcommand{\bl}{\mbox{\boldmath$l$}}
\newcommand{\pbpi}{\,^+\!\mbox{\boldmath$\pi$}}
\newcommand{\mbpi}{\,^-\!\mbox{\boldmath$\pi$}}
\newcommand{\pmbpi}{\,^{\pm}\!\mbox{\boldmath$\pi$}}
\renewcommand{\d}{{\rm d}}
\newcommand{\bpi}{\mbox{\boldmath$\pi$}}
\newcommand{\ppi}{\,^+\!\pi}
\newcommand{\ppisig}{\,^+\!\pi_{\sigma^2}}
\newcommand{\mpi}{\,^-\!\pi}
\newcommand{\mpisig}{\,^-\!\pi_{\sigma^2}}
\newcommand{\pOmega}{\,^+\!\Omega}
\newcommand{\mOmega}{\,^-\!\Omega}
\newcommand{\pmOmega}{\,^{\pm}\!\Omega}
\newcommand{\pv}{\,^+\!v}
\newcommand{\mv}{\,^-\!v}
\newcommand{\pmv}{\,^{\pm}\!v}
\newcommand{\pmSig}{\,^{\pm}\!\Sigma}
\newcommand{\ptau}{\,^+\!\tau}
\newcommand{\mtau}{\,^-\!\tau}
\newcommand{\Om}{\Omega^{\rm T}}
\newcommand{\pmR}{\,^{\pm}\!R}
\newcommand{\mR}{\,^-\!R}
\newcommand{\pR}{\,^+\!R}
\newcommand{\Rsig}{R_{\sigma^2}}
\newcommand{\pmRsig}{\,^{\pm}\!R_{\sigma^2}}
\newcommand{\pmcalN}{\,^{\pm}\!{\cal N}}
\newcommand{\pcalN}{\,^{+}\!{\cal N}}
\newcommand{\mcalN}{\,^{-}\!{\cal N}}
\newcommand{\pmm}{\,^{\pm}\!m}
\newcommand{\plm}{\,^+\!m}
\newcommand{\plmsig}{\,^+\!m_{\sigma^2}}
\newcommand{\mim}{\,^-\!m}
\newcommand{\mimsig}{\,^-\!m_{\sigma^2}}

\newcommand{\pmbphi}{\,^{\pm}\!\mbox{\boldmath$\phi$}}
\newcommand{\bvarphi}{\mbox{\boldmath$\varphi$}}
\newcommand{\bpsi}{\mbox{\boldmath$\psi$}}
\newcommand{\pmbSig}{\,^{\pm}\!\mbox{\boldmath$\Sigma$}}
\newcommand{\pmeta}{\,^{\pm}\!\eta}
\newcommand{\peta}{\,^+\!\eta}
\newcommand{\meta}{\,^-\!\eta}
\newcommand{\pmtheta}{\,^{\pm}\!\theta}
\newcommand{\pmtausig}{\,^{\pm}\!\tau_{\sigma^2}}
\newcommand{\pmpisig}{\,^{\pm}\!\pi_{\sigma^2}}
\newcommand{\pmzeta}{\,^{\pm}\!\zeta}
\newcommand{\pzeta}{\,^+\!\zeta}
\newcommand{\mzeta}{\,^-\!\zeta}
\newcommand{\palpha}{\,^+\!\alpha}
\newcommand{\pmbeta}{\,^{\pm}\!\beta}
\newcommand{\pmchi}{\,^{\pm}\!\chi}
\newcommand{\pchi}{\,^+\!\chi}
\newcommand{\mchi}{\,^-\!\chi}
\newcommand{\pmbnsig}{\,^{\pm}\!\mbox{\boldmath$n$}_{\sigma^2}}
\newcommand{\pmbeonesig}{\,^{\pm}\!\mbox{\boldmath$e$}_{\!1\sigma^2}}
\newcommand{\pmbeone}{\,^{\pm}\!\mbox{\boldmath$e$}_1}
\newcommand{\pmbetwo}{\,^{\pm}\!\mbox{\boldmath$e$}_2}
\newcommand{\pmbetwosig}{\,^{\pm}\!\mbox{\boldmath$e$}_{2\sigma^2}}
\newcommand{\pmbm}{\,^{\pm}\!\mbox{\boldmath$m$}}
\newcommand{\pmbv}{\,^{\pm}\!\mbox{\boldmath$v$}}
\newcommand{\pmmsig}{\,^{\pm}\!m_{\sigma^2}}
\newcommand{\pvsig}{\,^{+}\!v_{\sigma^2}}
\newcommand{\mvsig}{\,^{-}\!v_{\sigma^2}}
\newcommand{\tausig}{\tau_{\sigma^2}}
\newcommand{\pmzetasig}{\,^{\pm}\!\zeta_{\sigma^2}}
\newcommand{\pmetasig}{\,^{\pm}\!\eta_{\sigma^2}}
\newcommand{\pmchisig}{\,^{\pm}\!\chi_{\sigma^2}}
\newcommand{\pmbpisig}{\,^{\pm}\!\mbox{\boldmath$\pi$}_{\sigma^2}}
\renewcommand{\sinh}{{\rm sh}}
\renewcommand{\cosh}{{\rm ch}}

\begin{document}

\title{Barbero-Immirzi parameter in Regge calculus}
\author{V.M. Khatsymovsky \\
 {\em Budker Institute of Nuclear Physics} \\ {\em
 Novosibirsk,
 630090,
 Russia}
\\ {\em E-mail address: khatsym@inp.nsk.su}}
\date{}
\maketitle
\begin{abstract}
We consider Regge calculus in the representation in terms of area tensors and self-
and antiselfdual connections generalised to the case of Holst action that is standard
Einstein action in the tetrad-connection variables plus topological (on equations of
motion for connections) term with coefficient $1/\gamma$, $\gamma$ is Barbero-Immirzi
parameter. The quantum measure is shown to exponentially decrease with areas with
typical cut-off scales $4\pi G$ and $4\pi G\gamma$ in spacelike and timelike regions,
respectively ($G$ is the Newton constant).
\end{abstract}


\newpage

The formal nonrenormalisability of quantum version of general relativity (GR) may
cause us to try to find alternatives to the continuum description of underlying
spacetime structure. An example of such alternative description is given by Regge
calculus (RC) suggested in 1961 \cite{Regge}. It is the exact GR developed in the
piecewise flat spacetime which is a particular case of general Riemannian spacetime
\cite{Fried}. In its turn, the general Riemannian spacetime can be considered as a
limiting case of the piecewise flat spacetime \cite{Fein}. Any piecewise flat
spacetime is simlicial one: it can be represented as a collection of flat
4-dimensional {\it simplices}(tetrahedrons), and its geometry is completely specified
by a countable number of freely chosen lengths of all edges (or 1-simplices). Thus, RC
implies a {\it discrete} description {\it alternative} to the usual continuum one. For
a review of RC and alternative discrete gravity approaches see, e. g., \cite{RegWil}.

Since fully discrete theory such as RC does not possess a continuous coordinate
playing the role of time, the canonical Hamiltonian formalism and operator
quantization are not immediately applicable to it. However, the  functional integral
approach remains most universal. The functional integral measure in RC was considered
in \cite{HamWil1,HamWil2}. Our strategy (briefly reviewed in \cite{Kha0}) is based on
the requirement for the full discrete measure to result in the canonical Hamiltonian
functional integral measure with time $t$, with some coordinate chosen as $t$ and made
continuous.

Since this strategy implies intermediate use of canonical Hamiltonian measure, it is
of importance that we could perform continuous time limit in a nonsingular way.
Meanwhile, this limit implies infinitely flattened in some direction simplices, and
description of these objects is singular if made in terms of the edge lengths only.

The way to avoid singularities is to extend the set of variables via adding the new
ones having the sense of angles and considered as independent variables. Such
variables are the finite rotation matrices which are the discrete analogs of the
connections in the continuum GR. The situation considered is analogous to that one
occurred when recasting the Einstein action in the Hilbert-Palatini form. We consider
more general action which differs from the Hilbert-Palatini one by adding term which
is topological one on the equations of motion for the connection and thus leads to the
same Einstein action. Namely, we consider action introduced by Holst \cite{Holst}. He
has shown that his action leads to the Hamiltonian formalism by Barbero \cite{Barb}
which modifies Ashtekar formalism (see, e.g., review \cite{Ash}) to the case of real
variables. It also incorporates definition by Immirzi \cite{Imm} who extended
definition of \cite{Barb} to a whole family of quantum theories specified by parameter
$\gamma$ called recently Barbero-Immirzi parameter. The considered action reads
\begin{equation}                                                                    
\label{S-HilPal} \int{R\sqrt{g}{\rm d}^4x} ~~ \stackrel{\omega^{ab}_{\lambda} =
\omega^{ab}_{\lambda} (\{e^a_{\lambda}\}) }{\Leftarrow = = = = = =} ~~ {1\over
4}\int{(\epsilon_{abcd}e^a_{\lambda}e^b_{\mu} + {2\over \gamma}e_{\lambda c}e_{\mu d
})\epsilon^{\lambda\mu\nu\rho}
[\partial_{\nu}+\omega_{\nu},\partial_{\rho}+\omega_{\rho}]^{cd}{\rm d}^4x},
\end{equation}

\noindent where the tetrad $e^a_{\lambda}$ and connection $\omega^{ab}_{\lambda}$ =
$-\omega^{ba}_{\lambda}$ are independent variables, the RHS being reduced to LHS in
terms of $g_{\lambda\mu}$ = $e^a_{\lambda}e_{a\mu}$ if we substitute for
$\omega^{ab}_{\lambda}$ solution of the equations of motion for these variables in
terms of $e^a_{\lambda}$. The Latin indices $a$, $b$, $c$, ... are vector ones with
respect to the local Minkowskian frames introduced at each point $x$.

Now in RC the Einstein action in the LHS of (\ref{S-HilPal}) becomes the Regge action,
\begin{equation}                                                                    
\label{S-Regge} 2\sum_{\sigma^2}{\alpha_{\sigma^2}|\sigma^2|},
\end{equation}

\noindent where $|\sigma^2|$ is the area of a triangle (the 2-simplex) $\sigma^2$,
$\alpha_{\sigma^2}$ is the angle defect on this triangle, and summation run over all
the 2-simplices $\sigma^2$. The discrete analogs of the tetrad and connection, edge
vectors and finite rotation matrices, were first considered in \cite{Fro}. The local
inertial frames live in the 4-simplices. The analogs of the connection are defined on
the 3-simplices $\sigma^3$ and are the matrices $\Omega_{\sigma^3}$ connecting the
frames of the pairs of the 4-simplices $\sigma^4$ sharing the 3-faces $\sigma^3$.
These matrices are the finite SO(3,1) rotations in the Minkowskian case in contrast
with the continuum connections $\omega^{ab}_{\lambda}$ which are the elements of the
Lee algebra so(3,1) of this group. This definition includes pointing out the direction
in which the connection $\Omega_{\sigma^3}$ acts (and, correspondingly, the opposite
direction, in which the $\Omega^{-1}_{\sigma^3}$ = $\Omega^{\rm T}_{\sigma^3}$ acts).
That is, the connections $\Omega$ are defined on the {\it oriented} 3-simplices
$\sigma^3$. We have suggested self-dual formulation of RC \cite{Kha1} which easily
modifies from Euclidean to Minkowskian case and to include also topological term
considered here. Instead of RHS of (\ref{S-HilPal}) we write
\begin{eqnarray}                                                                    
\label{S-RegCon}%
S(v,\Omega) & = & \sum_{\sigma^2}{\left ( 1 + {i \over \gamma}
\right ) \sqrt{2\pvsig\circ\!\pvsig} \arcsin{\pvsig\circ
R_{\sigma^2}(\{\!\pOmega\})\over
\displaystyle\sqrt{2\pvsig\circ\!\pvsig}}}\nonumber\\ & & + \left
( 1 - {i \over \gamma} \right ) \sqrt{2\mvsig\circ\!\mvsig}
\arcsin{\mvsig\circ R_{\sigma^2}(\{\!\mOmega\})\over
\displaystyle\sqrt{2\mvsig\circ\!\mvsig}}
\end{eqnarray}

\noindent where we have defined $A\circ B$ = ${1\over 2}A^{ab}B_{ab}$ for the two
tensors $A$, $B$; $\{\dots\}$ means "the set of \dots "; $v_{\sigma^2}$ is the dual
tensor of the triangle $\sigma^2$ in terms of the vectors of its edges $l^a_i$,
\begin{equation}                                                                    
\label{v=ll} v_{\sigma^2ab} = {1\over 2}\epsilon_{abcd}l^c_1l^d_2
\end{equation}

\noindent (in some 4-simplex frame containing $\sigma^2$). The curvature matrix
$R_{\sigma^2}$ on the 2-simplex $\sigma^2$ is the path ordered product of the
connections $\Omega^{\pm 1}_{\sigma^3}$ on the 3-simplices $\sigma^3$ sharing
$\sigma^2$ along the contour enclosing $\sigma^2$ once and contained in the
4-simplices sharing $\sigma^2$,
\begin{equation}                                                                    
\label{R-Omega} R_{\sigma^2} = \prod_{\sigma^3\supset\sigma^2}{\Omega^{\pm
1}_{\sigma^3}}.
\end{equation}

\noindent The $^{\pm}\!(\dots)$-notations (self- and antiselfdual parts) are as
follows. For SO(3,1) matrix
$$ \Omega = \exp {(\varphi^kE^a_{kb} + \psi^kL^a_{kb})} $$
with generators
$$ E_{kab} = -\epsilon_{kab}, ~~~ L_{kab} = g_{ka}g_{0b} - g_{0a}g_{kb}~~~(g_{ab} =
{\rm diag}(-1,1,1,1), \epsilon_{123} = +1 ) $$
we define
$$ \pmSig_{kab} = -\epsilon_{kab} \pm i(g_{ak}g_{0b} - g_{a0}g_{kb}) $$
so that
$$ \pmSig^a_{kb} \pmSig^b_{lc} = - \delta_{kl} \delta^a_c + \epsilon_{kl}^{~~m}
\pmSig^a_{mc} $$
and then
$$ \Omega = \pOmega \mOmega, ~~~ \pmOmega = \exp \left ({\varphi^k \mp i\psi^k \over
2}\pmSig^a_{kb} \right ). $$
Area tensor $v$ splits additively,
$$ v^{ab} = \pv^{ab} + \mv^{ab}, ~~~ \pmv^{ab} = {1\over 2} v^{ab} \pm {i\over 4}
\epsilon^{ab}_{~~cd}v^{cd} ~~~ ( \epsilon^{0123} = +1) $$
so that
$$ *v^{ab} \equiv {1\over 2} \epsilon^{ab}_{~~cd}\pmv^{cd} = \mp i\pmv^{ab}. $$
In particular,
$$ 2\pmv \circ \pmv = v \circ v \pm iv*v. $$
The $^{\pm}$-parts map to 3d vectors $ \pmbv $,
$$ \pmv_{ab} \equiv {1\over 2}\pmv^k\pmSig_{kab}, ~~~ 2\pmv_k = - \epsilon_{klm}
v^{lm} \pm i(v_{k0} -v_{0k}). $$
In particular,
$$ \pmbv^2 = 2\pmv \circ \pmv. $$
For $v_{\sigma^2}$ given by (\ref{v=ll}) the $\pmbv_{\sigma^2}^2$ is $(-4)$ times
square of (real for spacelike $\sigma^2$) area.

Classically, we can write equations of motion for $\pmOmega_{\sigma^3}$, that is, for
the corresponding parameters $\bvarphi$, $\bpsi$. This results in equations for
$\bvarphi - i\bpsi$ and $\bvarphi + i\bpsi$ separately for ${}^+$- and ${}^-$-parts as
for holomorphic functions. Take ${}^+$-part. Dependence on $\pOmega_{\sigma^3}$ is due
to contributions from the faces $\sigma^2$ of the tetrahedron $\sigma^3$; take certain
such face with tensor $\pvsig \equiv \pv$ . Consider $U \equiv \pOmega_{\sigma^3}$ as
a'priori arbitrary $4\times 4$ matrix but add Lagrange multiplier terms taking into
account orthogonality of $U$ and its self-duality (equivalent to self-duality of its
antisymmetric part). The dependence on $U$ in action is
\begin{eqnarray}                                                                    
S & \propto & \sqrt{2\pv \circ \pv}\arcsin {\pv \circ (\Gamma_1 U \Gamma_2)\over
\sqrt{2\pv \circ \pv}} + (U^{\rm T} \circ U - 1) \circ \lambda + (*U + iU) \circ \mu,
\nonumber
\\ & &
\Gamma_1 U \Gamma_2 = R(\{\!\pOmega\}) \mbox{ or } R^{\rm T}(\{\!\pOmega\})
\end{eqnarray}

\noindent Here symmetric $\lambda$ and antisymmetric $\mu$ matrices are Lagrange
multipliers. Let us form combination of the equations of motion
\begin{equation}                                                                    
i\epsilon_{fg}{}^{ab}U_a{}^c{\!\!\!\partial S\over \partial U^{bc}} + U_f{}^c {
\!\!\!\partial S \over \partial U^{gc}} - U^{~c}_g { \!\!\!\partial S \over \partial
U^{fc}} = 0
\end{equation}

\noindent (in fact, ${}^+$-part of $U^{\rm T} \partial S / \partial U$), thereby
$\lambda$- and $\mu$-terms are cancelled. We are left with the ${}^+$-part of some
product of ${}^+$-matrices which coincides with this product itself (antisymmetrised),
\begin{equation}\label{eqs-motion}                                                  
\Gamma^{\rm T}_1\frac{R(\{\!\pOmega\})\pv + \pv R^{\rm T}(\{\!\pOmega\})}{\cos
\palpha}\Gamma_1 ~~~ \left( \palpha = \arcsin \frac{\pv \circ
R(\{\!\pOmega\})}{\sqrt{2\pv \circ \pv}} \right).
\end{equation}

\noindent Take as $\{\Omega\}$ the set of compatible with edge lengths connections so
that $\Rsig (\{\Omega\})$ really rotates around $\sigma^2$ by the defect angle. In
correspondence with Minkowskian metric signature, there are two types of the
rotations.

\noindent i). Euclidean rotation around timelike area. In certain frame we have for
triangle of area $1/2$ (in module),
\begin{eqnarray}
v_{ab} = \left( \begin{array}{cccc}
0 & 0 & 0 & 0 \\
0 & 0 & 0 & 0 \\
0 & 0 & 0 & -1 \\
0 & 0 & 1 & 0
\end{array} \right),
~~~ 2\pmv_{ab} = \left( \begin{array}{cccc}
0 & \pm i & 0 & 0 \\
\mp i & 0 & 0 & 0 \\
0 & 0 & 0 & -1 \\
0 & 0 & 1 & 0
\end{array} \right), ~~~ \pmbv = (1, 0, 0),\nonumber\\
\pmR^{ab} = \left( \begin{array}{cccc}
-\cos\frac{\varphi}{2} & \mp i\sin\frac{\varphi}{2} & 0 & 0 \\
\pm i\sin\frac{\varphi}{2} & \cos\frac{\varphi}{2} & 0 & 0 \\
0 & 0 & \cos\frac{\varphi}{2} & -\sin\frac{\varphi}{2} \\
0 & 0 & \sin\frac{\varphi}{2} & \cos\frac{\varphi}{2}
\end{array} \right), ~~~ \sqrt{\pmbv^2} \arcsin \frac{\pmv \circ \pmR}{\sqrt{\pmbv^2}}
= \frac{\varphi}{2}.                                                                
\end{eqnarray}

\noindent ii). Rotation around spacelike area, i.e. Lorentz boost,
\begin{eqnarray}
v_{ab} = \left( \begin{array}{cccc}
0 & -1 & 0 & 0 \\
1 & 0 & 0 & 0 \\
0 & 0 & 0 & 0 \\
0 & 0 & 0 & 0
\end{array} \right),
~~~ 2\pmv_{ab} = \left( \begin{array}{cccc}
0 & -1 & 0 & 0 \\
1 & 0 & 0 & 0 \\
0 & 0 & 0 & \mp i \\
0 & 0 & \pm i & 0
\end{array} \right), ~~~ \pmbv = (\pm i, 0, 0),\nonumber\\
\pmR^{ab} = \left( \begin{array}{cccc}
-\cosh\frac{\psi}{2} & -\sinh\frac{\psi}{2} & 0 & 0 \\
\sinh\frac{\psi}{2} & \cosh\frac{\psi}{2} & 0 & 0 \\
0 & 0 & \cosh\frac{\psi}{2} & \pm i\sinh\frac{\psi}{2} \\
0 & 0 & \mp i\sinh\frac{\psi}{2} & \cosh\frac{\psi}{2}
\end{array} \right), ~~~ \sqrt{\pmbv^2} \arcsin \frac{\pmv \circ \pmR}{\sqrt{\pmbv^2}}
= \frac{\psi}{2}.                                                                  
\end{eqnarray}

\noindent In both cases the ${}^+$- and ${}^-$-parts contribute the same half of the
action (\ref{S-Regge}) and reproduce it while topological term is cancelled. What is
important, in (\ref{eqs-motion}) we have $R(\{\!\pOmega\})\pv + \pv R^{\rm
T}(\{\!\pOmega\}) = 2\pv\cos\!\palpha$ and proportional to
\begin{equation}                                                                   
\Gamma^{\rm T}_1\pvsig\Gamma_1
\end{equation}

\noindent contribution of the given 2-face $\sigma^2$ of $\sigma^3$ into equations of
motion for $\pOmega_{\sigma^3}$. Here matrices $\Gamma$ serve to transform $\pvsig$ to
the same local frame for all faces of the given $\sigma^3$. As a result, we get simply
the closure condition for the surface of $\sigma^3$ fulfilled identically by
construction of the manifold. Thus equations of motion for connections really lead to
the original Regge action.

Can we fix full discrete measure from requirement to result in the Hamiltonian path
integral measure in the continuous time limit whatever coordinate is chosen as time?
This strategy has solution in 3 dimensions. A specific feature of the 3D case
important here is commutativity of the dynamical constraints in the (continuous time)
Hamiltonian formulation leading to a simple form of the functional integral. In 3D
case egde vectors $\bl_{\sigma^1}$ are considered instead of area tensors
$v_{\sigma^2}$. Edge vectors are independent of each other, at least locally. In 4
dimensions, the variables $v_{\sigma^2}$ are not independent but obey a set of
(bilinear) {\it intersection relations}. For example, tensors of the two triangles
$\sigma^2_1$, $\sigma^2_2$ sharing an edge satisfy the relation
\begin{equation}                                                                   
\label{v*v}%
\epsilon_{abcd}v^{ab}_{\sigma^2_1}v^{cd}_{\sigma^2_2} = 0.
\end{equation}

\noindent These purely geometrical relations define a hypersurface in the
configuration superspace of the formally independent area tensors. The idea is to
construct quantum measure first in this superspace. At the second stage the measure
should be projected onto this hypersurface: geometrical relations of the type
(\ref{v*v}) are taken into account by inserting certain $\delta$-function-like factors
in the measure. Note that the RC with formally independent (scalar) areas has been
considered in the literature \cite{RegWil,BarRocWil}.

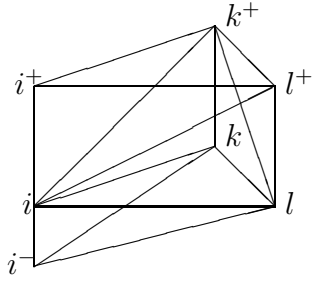
\begin{figure}\unitlength 0.20mm
\begin{picture}(200,200)(-150,20)
\put (150,20){\line(0,1){120}} \put (150,20){\line(3,2){120}} \put
(150,20){\line(4,1){160}} \put (150,60){\line(1,1){120}} \put
(150,60){\line(2,1){160}} \put (150,60){\line(3,1){120}} \put
(150,60){\line(1,0){160}} \put (150,140){\line(3,1){120}} \put
(150,140){\line(1,0){160}} \put (270,100){\line(0,1){80}} \put
(270,100){\line(1,-1){40}} \put (270,180){\line(1,-1){40}} \put
(270,180){\line(1,-3){40}} \put (310,60){\line(0,1){80}} \put (310,135){$~l^{+}$} \put
(270,180){$~k^{+}$} \put (137,135){$i^{+}$} \put (142,55){$i$} \put (270,100){$~k$}
\put (310,55){$~l$} \put (132,15){$i^{-}$}
\end{picture}
\renewcommand{\baselinestretch}{1.0}
\caption{Fragment of the $t$-like 3-prism.}\label{3prism}
\end{figure}

The theory with formally independent area tensors can be called area tensor RC. The
completely discrete quantum measure reads
\begin{eqnarray}                                                                    
\label{VEV2}%
<\Psi (\{\pi\},\{\Omega\})> & = & \int{\Psi (\{\pi\}, \{\Omega\})\exp{\left \{i\!
\sum_{\stackrel{t-{\rm like}}{\sigma^2}}{\left [\left (1 + {i\over \gamma}\right
)\ptau _{\sigma^2}\circ
R_{\sigma^2}(\{\!\pOmega\})\right .}\right .}}\nonumber\\
 & &\left. \left. \hspace{-40mm} + \left (1 - {i\over \gamma}\right
)\mtau _{\sigma^2}\circ R_{\sigma^2}(\{\!\mOmega\}) \right ] + i
\!\sum_{\stackrel{\stackrel{\rm not}{t-{\rm like}}}{\sigma^2}} \left [ \left (1 +
{i\over \gamma} \right ) \ppi_{\sigma^2}\circ
R_{\sigma^2}(\{\!\pOmega\}) \right. \right. \nonumber\\
 & &\left. \left. \hspace{-40mm} + \left (1 - {i\over \gamma}\right
) \mpi_{\sigma^2}\circ R_{\sigma^2}(\{\!\mOmega\}) \right ] \right \}
\prod_{\stackrel{\stackrel{\rm
 not}{t-{\rm like}}}{\sigma^2}}{\rm d}^6
\pi_{\sigma^2}\prod_{\sigma^3}{{\cal D}\Omega_{\sigma^3}} \nonumber\\
  & \equiv &
\int{\Psi (\{\pi\},\{\Omega\}){\rm d} \mu_{\rm area}(\{\pi\},\{\Omega\})}
\end{eqnarray}

\noindent where ${\cal D}\Omega_{\sigma^3}$ is the Haar measure on the group SO(3,1)
of connection matrices $\Omega_{\sigma^3}$. Appearance of some set ${\cal F}$ of
triangles $\sigma^2$ integration over area tensors of which is omitted (denoted as
"$t$-like" in (\ref{VEV2}))is connected with that integration over {\it all} area
tensors is generally infinite, in particular, when normalizing measure (finding
$<1>$). Indeed, different $R_{\sigma^2}$ for $\sigma^2$ meeting at a given link
$\sigma^1$ are connected by Bianchi identities \cite{Regge}. Therefore the product of
$\delta^6(R_{\sigma^2} - R^{\rm T}_{\sigma^2})$ for all these $\sigma^2$ which follows
upon integration over area tensors for these $\sigma^2$ contains singularity of the
type of $\delta$-function squared. To avoid this singularity we should confine
ourselves by only integration over area tensors on those $\sigma^2$ on which
$R_{\sigma^2}$ are independent. The complement ${\cal F}$ to this set of $\sigma^2$
are those $\sigma^2$ on which $R_{\sigma^2}$ are dependent, that is, expressible by
the Bianchi identities in terms of independent $R_{\sigma^2}$. Let us adopt regular
way of constructing 4D simplicial structure of the 3D simplicial geometries (leaves)
of the same structure. A $n$-simplex $\sigma^n$ is denoted by the set of its $n + 1$
vertices in round brackets (unordered sequence), $(A_1A_2...)$. The $i$, $k$, $l$, ...
are vertices of the current leaf, $i^+$, $k^+$, $l^+$, ... and $i^-$, $k^-$, $l^-$,
... are corresponding vertices of the nearest future and past in $t$ leaves. Or we
shall speak of the "upper" and "lower" leaves, respectively. See fig.\ref{3prism}.
Each vertex is connected by links (edges) with its $\pm$-images. These links (of the
type of $(ii^+)$, $(ii^-)$) will be called $t$-{\it like} ones (do not mix with the
term "timelike" which is reserved for the local frame components). The {\it leaf}
links $(ik)$ are completely contained in the 3D leaf. There may be {\it diagonal}
links $(ik^+)$, $(ik^-)$ connecting a vertex with the $\pm$-images of its neighbors.
We call arbitrary simplex $t$-{\it like} one if it has $t$-like edge, the {\it leaf}
one if it is completely contained in the 3D leaf and {\it diagonal} one in other
cases. It can be seen that the set of the $t$-like triangles is fit for the role of
the above set ${\cal F}$. In the case of general 4D simplicial structure we can deduce
that the set ${\cal F}$ of the triangles with the Bianchi-dependent curvatures pick
out some one-dimensional field of links. We can take this field as definition of the
coordinate $t$ direction so that ${\cal F}$ be just the set of the $t$-like triangles.
Also existence of the set ${\cal F}$ naturally fits our above requirement that
limiting form of the full discrete measure (when any one of the coordinates, not
necessarily $t$, is made continuous) should coincide with Hamiltonian path integral.
Namely, in the Hamiltonian formalism absence of integration over area tensors of
triangles which pick out some coordinate $t$ ($t$-like ones) corresponds to some gauge
fixing.

There is the invariant (Haar) measure ${\cal D}\Omega$ in (\ref{VEV2}) which looks
natural from symmetry considerations. But it also arises from the formal point of view
within our strategy. If one makes a coordinate $t$ continuous and takes it as a time
he finds kinetic terms of the type $\pi_{\sigma^2}\circ\Omega^{\rm T
}_{\sigma^2}\dot{\Omega}_{\sigma^2}$ (here $\Omega_{\sigma^2}$ serves to parameterize
limiting form of $\Omega_{\sigma^3}$ with $\sigma^3$ filling up infinitesimal $t$-like
prism with the base $\sigma^2$). Then standard Hamiltonian path integral just has
${\cal D}\Omega_{\sigma^2}$ as a measure. To reproduce the latter in the continuous
$t$ limit the full discrete measure should also include Haar measure ${\cal
D}\Omega_{\sigma^3}$.

One else specific feature of the quantum measure is the absence of the inverse
trigonometric function 'arcsin' in the exponential, whereas the Regge action
($\ref{S-RegCon}$) contains such functions. This is connected with using the canonical
quantization at the intermediate stage of derivation: in gravity this quantization is
completely defined by the constraints, the latter being equivalent to those ones
without $\arcsin$ (in some sense on-shell).

Consider averaging functions of only area tensors $\pi_{\sigma^2}$. By the properties
of invariant measure, integrations over $\prod{\cal D}\Omega_{\sigma^3}$ in
(\ref{VEV2}) reduce to integrations over $\prod{\cal D}R_{\sigma^2}$ with independent
$R_{\sigma^2}$ (i.e. at not $t$-like $\sigma^2$) and some number of connections
$\prod{\cal D}\Omega_{\sigma^3}$ which we can call gauge ones. The expectation value
of any field monomial, $<\pi^{a_1b_1}_{\sigma^2_1}...\pi^{a_nb_n}_{\sigma^2_n}>$
reduces to the (derivatives of) $\delta$-functions $\delta^3 ((\pR_{\sigma^2_i} -
\pR^{\rm T}_{\sigma^2_i}) + (\mR_{\sigma^2_i} - \mR^{\rm T}_{\sigma^2_i}))$ $\propto$
$\delta^3 (\Re (\pR_{\sigma^2_i} - \pR^{\rm T}_{\sigma^2_i}))$ and $\delta^3
(i[(\pR_{\sigma^2_i} - \pR^{\rm T}_{\sigma^2_i}) - (\mR_{\sigma^2_i} - \mR^{\rm
T}_{\sigma^2_i})])$ $\propto$ $\delta^3 (\Im (\pR_{\sigma^2_i} - \pR^{\rm
T}_{\sigma^2_i}))$ which are then integrated out over ${\cal D}R_{\sigma^2_i}$ giving
finite nonzero answer. This finiteness means that the result of integrations over
connections should be function of $\pi_{\sigma^2}$ sufficiently rapidly decreasing at
infinity. If we try to get monotonic exponent from oscillating one by moving
integration contour over curvature to complex plane, we should eventually have
decreasing exponent. Consider typical integral
\begin{equation}                                                                   
\label{basic} \int \exp \left [ i\left (1 + {i\over \gamma} \right ) \ppi \circ \pR +
i\left (1 - {i\over \gamma}\right ) \mpi \circ \mR \right ] {\cal D}R.
\end{equation}

\noindent The $R$ is parameterized by 6 variables $\bvarphi$, $\bpsi$. Let us deform
integration contour over $\bvarphi$, $\bpsi$ in the complex space ${\rm C}^6$ in the
following way.
\begin{eqnarray}\label{transform}                                                  
{1\over 2}\left(\bvarphi \mp i\bpsi\right) & \Longrightarrow & \left [\displaystyle
{-i\left ( 1 \pm {i\over\gamma} \right ) \pmbpi \over \sqrt{-\left ( 1 \pm
{i\over\gamma} \right )^2 \pmbpi^2} } \cosh \pmzeta \right. \nonumber\\
 &  & \left. + i(\sinh \pmzeta)(\pmbeone \cos \pmchi + \pmbetwo
\sin \pmchi) \right ] \left ( {\pi\over 2} + i\pmeta \right ).
\end{eqnarray}

\noindent Here $\pmbeone^2 = 1 = \pmbetwo^2$, $\pmbeone\cdot\pmbetwo = 0$,
$\pmbeone\cdot\pmbpi = 0 = \pmbetwo\cdot\pmbpi$ (that is, orthonormal to $\pmbpi$ the
double). The $\sqrt{z}$ is defined in the plane ${\rm C}$ with cut off along negative
real half-axis $\Im z = 0, \Re z \leq 0$ so that $\sqrt{1}$ = 1. Integral becomes that
over $\pmeta,\pmzeta,\pmchi$,
\begin{eqnarray}                                                                   
&&\hspace{-15mm}\left (4\pi\right )^{-2}\int \exp \left [ -\sqrt{-\left ( 1 + {i\over
\gamma} \right )^2\pbpi^2} \cosh \!\peta \,\cosh \!\pzeta -\sqrt{-\left ( 1 - {i\over
\gamma} \right )^2\mbpi^2} \cosh \!\meta \,\cosh \!\mzeta \right ]\nonumber\\&& \cdot
\cosh^2 \!\peta\, \d \!\peta \,\d \cosh \!\pzeta \,\d \!\pchi \cdot \,\cosh^2
\!\meta\, \d \!\meta \,\d
\cosh \!\mzeta \,\d \!\mchi \nonumber\\
& = & {K_1 \left[\sqrt{-\left ( 1 + {i\over \gamma} \right )^2\pbpi^2 }\right]\over
\sqrt{-\left ( 1 + {i\over \gamma} \right )^2\pbpi^2 }} \cdot {K_1 \left[\sqrt{-\left
( 1 - {i\over \gamma} \right )^2\mbpi^2 }\right]\over \sqrt{-\left ( 1 - {i\over
\gamma} \right )^2\mbpi^2 }}.
\end{eqnarray}

The $K_1$ is the modified Bessel function. Transform (\ref{transform}) has simple
sense if divided into two stages. i) Make $\bpsi$ imaginary so that $(\bvarphi \mp
i\bpsi)/2$ become independent 3-vector real variables $\pmbphi$. Then it becomes
possible to split the measure
\begin{equation}                                                                   
{\cal D}R = {\cal D}\pR {\cal D}\mR.
\end{equation}

\noindent ii) Transform spherical components of $\pmbphi$: move $\sqrt{\pmbphi^2}$ to
$\pi/2 + i\pmeta$, $-\infty < \pmeta < +\infty$, move the azimuthal angle $\pmtheta$
of $\pmbphi$ w.r.t. $\pmbpi$ to $i\pmzeta$, $0 \leq \pmzeta < +\infty$, the polar
angle $\pmchi$ remaining the same.

On physical hypersurface (\ref{v=ll}) $\pbpi^2 = \mbpi^2 \equiv \bpi^2$. For spacelike
(i.e. usual) areas $\bpi^2 < 0$ and we have
\begin{equation}                                                                   
\left ( 1 + {1\over \gamma^2} \right )^{-1}(-\bpi^2)^{-1}K_1 \left [ \left ( 1 +
{i\over \gamma} \right )\sqrt{-\bpi^2} \right ]K_1 \left [ \left ( 1 - {i\over \gamma}
\right )\sqrt{-\bpi^2} \right ]
\end{equation}

\noindent and for timelike areas ($\bpi^2 > 0$)
\begin{equation}                                                                   
\left ( 1 + {1\over \gamma^2} \right )^{-1}(\bpi^2)^{-1}K_1 \left [ \left ( {1\over
\gamma} - i \right )\sqrt{\bpi^2} \right ]K_1 \left [ \left ( {1\over \gamma} + i
\right )\sqrt{\bpi^2} \right ].
\end{equation}

\noindent Both these have the same asymptotic form at $\bpi^2 \to 0$,
\begin{equation}                                                                   
\left ( 1 + {1\over \gamma^2} \right )^{-2}(\bpi^2)^{-2},
\end{equation}

\noindent but differ at $|\bpi^2 | \to \infty$,
\begin{equation}                                                                   
{\pi\over 2}\left ( 1 + {1\over \gamma^2} \right )^{-3/2}|\bpi^2|^{-3/2}\cdot \left\{
\begin{array}{rl} \exp \left(\displaystyle -2{{}\over {}}\!\sqrt{|\bpi^2|}
\right), & \mbox{at } \bpi^2 < 0, \\ \exp \left(\displaystyle -{2\over
\gamma}\sqrt{|\bpi^2|} \right), & \mbox{at } \bpi^2 >0.
\end{array} \right.
\end{equation}

\noindent In usual units (coefficient $(16\pi G)^{-1}$ at the action) and taking into
account that $|\bpi^2|^{-1/2}$ is twice $A$, module of the triangle area, we find
exponential decrease proportional to $\exp (-A(4\pi G)^{-1})$ and $\exp (-A(4\pi G
\gamma )^{-1})$ in the spacelike and timelike regions, respectively.

In reality $\tausig \neq 0$. The $\Rsig$ in the terms $\pmtausig \circ \pmRsig$ are by
Bianchi identities functions of $\{\Rsig | \sigma^2 \mbox{ not } t\mbox{-like}\}$. But
any given $\Rsig$ for not $t$-like $\sigma^2$ enters these terms only linearly (or
does not enter at all). Therefore for any given $\Rsig$ at not $t$-like $\sigma^2$ we
have integral of the type (\ref{basic}) but $\pmpisig$ should be replaced by some
matrix $\pmmsig$. Here $\pmmsig$ differs from $\pmpisig$ by a sum of (multiplied by
products of some $\pmRsig$) some $\pmtausig$ for the $\sigma^2$s forming lateral
surface of 3-prism considered below. Area tensor of one of the bases of this prism is
just this $\pmpisig$.

The $\pmmsig$ has not only antisymmetric but also trace part,
\begin{equation}                                                                   
\pmm = {1\over
2}\pmm_k \cdot \pmSig^k + {1\over 2}\pmm_0 \cdot 1, ~~~ 2\pmm \circ \pmm = \pmbm^2 +
\pmm_0^2.
\end{equation}

(Here notations $\pmm$, $\pmm_0$ do not mean $^{\pm}$-parts of anything.) The integral
(\ref{basic}) generalises to
\begin{eqnarray}                                                                   
\label{basic-m}&& \left (4\pi\right )^{-2}\int \exp \left [ \left (1 + {i\over \gamma}
\right ) \plm \circ \pR + \left (1 - {i\over \gamma}\right ) \mim \circ \mR \right ]
{\cal D}R\nonumber\\ & = & {K_1 \left[\sqrt{-\left ( 1 + {i\over \gamma} \right
)^2\cdot 2\plm \circ \plm }\right]\over \sqrt{-\left ( 1 + {i\over \gamma} \right
)^2\cdot 2\plm \circ \plm }} \cdot {K_1 \left[\sqrt{-\left ( 1 - {i\over \gamma}
\right )^2\cdot 2\mim \circ \mim }\right]\over \sqrt{-\left ( 1 - {i\over \gamma}
\right )^2\cdot 2\mim \circ \mim }}.
\end{eqnarray}

\noindent (Appropriate complex deformation of integration contours reads
\begin{eqnarray}\label{transform-m}                                                
{1\over 2}\left(\bvarphi \mp i\bpsi\right) & \Longrightarrow & \left [\displaystyle
{-i\left ( 1 \pm {i\over\gamma} \right ) \pmbm \over \sqrt{-\left ( 1 \pm
{i\over\gamma} \right )^2 \pmbm^2} } \cosh \pmzeta \right. \nonumber\\
 &  & \left. + i(\sinh \pmzeta)(\pmbeone \cos \pmchi + \pmbetwo
\sin \pmchi) \right ] \left ( {\pi\over 2} + i\pmeta - \pmbeta \right )
\end{eqnarray}

\noindent where
\begin{equation}                                                                   
\left (\cos \!\pmbeta ~ ; ~ \sin \!\pmbeta\right ) = {\left (\sqrt{-\left ( 1 \pm
{i\over \gamma}\right )^2\pmbm^2} \cosh \!\pmzeta ~ ; ~~~ -i\left ( 1 \pm {i\over
\gamma}\right )\pmm_0 \right ) \over \sqrt{-\left ( 1 \pm {i\over \gamma}\right
)^2\left (\pmbm^2\cosh^2 \!\pmzeta + \pmm^2_0\right )}}.
\end{equation}

\noindent ) The idea is to try to find some set of the 2-simplices ${\cal M}$ so that
exponential in (\ref{VEV2}) be representable in the form
\begin{eqnarray}\label{-mR-piR}                                                    
i\sum_{\sigma^2 \in {\cal M}} \left [\left (1 + {i\over \gamma}\right )\plmsig \circ
\Rsig (\{\!\pOmega\}) + \left (1 - {i\over \gamma}\right )\mimsig \circ \Rsig
(\{\!\mOmega\}) \right ]\nonumber \\ + i\sum_{\sigma^2 \not\in {\cal M}} \left [\left
(1 + {i\over \gamma}\right )\ppisig \circ \Rsig (\{\!\pOmega\}) + \left (1 - {i\over
\gamma}\right )\mpisig \circ \Rsig (\{\!\mOmega\}) \right ]
\end{eqnarray}

\noindent where $\pmmsig$ = $\pmpisig$ + (linear in $\{ \pmtausig \}$ terms). The set
$\{ \pmmsig \}$ depend on $\{ \pmtausig \}$ and on $\{ \pmRsig | \sigma^2 \not\in
{\cal M} \}$, but not on $\{ \pmRsig | \sigma^2 \in {\cal M} \}$. Then integrations
over $\{ \Rsig | \sigma^2 \in {\cal M} \}$ can be explicitly performed according to
eq. (\ref{basic-m}). For other $\{ \Rsig | \sigma^2 \not\in {\cal M} \}$ deformation
of integration contours according to eq. (\ref{transform}) is made. The result reads
\begin{eqnarray}\label{dN1}                                                        
& &\hspace{-10mm} \d \mu_{\rm area} \equiv \pcalN \mcalN
\prod_{\stackrel{\stackrel{\rm not}
{t-{\rm like}}}{\sigma^2}}{\rm d}^3 \pmbpi_{\sigma^2};~~~ 
\pmcalN = \int \left \{ \prod_{\sigma^2 \in {\cal M}}{K_1\left [\sqrt{-\left ( 1 \pm
{i\over \gamma} \right )^2 2\pmmsig \circ \pmmsig}\right ] \over \sqrt{-\left ( 1 \pm
{i\over \gamma} \right )^2 2\pmmsig \circ \pmmsig}} \right \}\nonumber\\
& & \hspace{-10mm} \cdot \exp \left [ -\sum_{\stackrel{\stackrel{\rm not} {t-{\rm
like}}}{\sigma^2 \not \in {\cal M}}} \sqrt{-\left (1 \pm {i\over \gamma}\right
)^2\pmbpisig^2 }\cosh \!\pmzetasig \cosh \!\pmetasig \right ]
\prod_{\stackrel{\stackrel{\rm not} {t-{\rm like}}}{\sigma^2 \not \in {\cal M}}}
\cosh^2 \!\pmetasig \d \pmetasig \d \cosh \! \pmzetasig \d \pmchisig
\end{eqnarray}

\noindent where $\{ \pmmsig | \sigma^2 \in {\cal M} \}$ depend on $\{ \pmetasig,
\pmzetasig, \pmchisig | \sigma^2 \not\in {\cal M} \}$ through $\Rsig$ parameterized by
these,
\begin{eqnarray}\label{rotated-R}                                                  
& & \hspace{-25mm} \pmRsig = -i\sinh \!\pmetasig + \pmbSig \cdot \!\pmbnsig \cosh
\!\pmetasig, \nonumber\\& & \hspace{-25mm} \pmbnsig = {-i\left (1 \pm {i\over
\gamma}\right )\pmbpisig \over \sqrt{-\left (1 \pm {i\over \gamma}\right )^2
\pmbpisig^2 }} \cosh \!\pmzetasig + i(\sinh \!\pmzetasig )(\pmbeonesig \cos
\!\pmchisig + \pmbetwosig \sin \!\pmchisig).
\end{eqnarray}

\noindent This looks as product of exponentially dumped at large areas multipliers
(note that $\Re \sqrt{z}$ $\!\geq\!$ 0 for our choice of the branch of function
$\sqrt{z}$ with cut along negative real half-axis in the complex plane of $z$ such
that $\sqrt{1}$ = 1).

To construct the set ${\cal M}$, note that due to the Bianchi identities dependence on
the matrix $\Rsig$ on the given leaf/diagonal triangle $\sigma^2$ in the exponential
of (\ref{VEV2}) comes from all the triangles constituting together with this
$\sigma^2$ a closed surface. This is surface of the $t$-like 3-prism, one base of
which is just the given $\sigma^2$, the lateral surface consists of $t$-like triangles
and goes to infinity. In practice, replace this infinity by some lowest (initial) leaf
where another base $\sigma^2_0$ is located the tensor of which $\pi_{\sigma^2_0}$ is
taken as boundary value (it is implied $\pi_{\sigma^2_0}$ = 0 above). Consider a
variety of such prisms with upper bases $\sigma^2$ placed in the uppest (final) leaf
such that any link in this leaf belongs to one and only one of these bases. That is,
lateral surfaces of different prisms do not have common triangles. Then the terms
$\pmmsig \circ \pmRsig$ in (\ref{-mR-piR}) represent contribution from these prisms,
${\cal M}$ being the set of their bases in the uppest leaf.

To really reduce the measure to such form, we should express the curvature matrices on
the $t$-like triangles in terms of those on the leaf/diagonal ones. The curvature on a
leaf/diagonal triangle $\sigma^2$ as product of $\Omega$s includes the two matrices
$\Omega$ on the $t$-like tetrahedrons $\sigma^3$ adjacent to $\sigma^2$ from above and
from below. Knowing curvatures on the set of leaf/diagonal triangles inside any t-like
3-prism allows to successively express matrix $\Omega$ on any $t$-like tetrahedron
inside the prism in terms of matrix $\Omega$ on the uppest $t$-like tetrahedron in
this prism taken as boundary value. Expressions for the considered curvatures look
like (fig.\ref{3prism})
\begin{eqnarray}                                                                   
& & \hspace{-18mm} \dots \dots \dots \dots \dots \dots \dots \dots \dots \dots \dots
\nonumber\\ R_{(ikl)} & = & \dots \Om_{(i^-ikl)} \dots \Omega_{(ik^+kl)} \dots
\nonumber\\ R_{(ik^+)} & = & \dots \Om_{(ik^+kl)} \dots \Omega_{(ik^+l^+l)} \dots \\
R_{(ik^+l^+)} & = & \dots \Om_{(ik^+l^+l)} \dots \Omega_{(i^+ik^+l^+)} \dots
\nonumber\\ & & \hspace{-18mm} \dots \dots \dots \dots \dots \dots \dots \dots \dots
\dots \dots \nonumber
\end{eqnarray}

\noindent The dots in expressions for $R$ mean matrices $\Omega$ on the leaf/diagonal
tetrahedrons which can be considered as gauge ones. We can step-by-step express
$\Omega_{(i^-ikl)}$ $\rightarrow$ $\Omega_{(ik^+kl)}$ $\rightarrow$
$\Omega_{(ik^+l^+l)}$ $\rightarrow$ $\Omega_{(i^+ik^+l^+)}$ $\rightarrow$ \dots where
the arrow means "in terms of". Knowing $\Omega$s on $t$-like tetrahedrons we can find
the curvatures on $t$-like triangles, the products of these $\Omega$s,
\begin{equation}                                                                   
R_{(i^+ikl)} = \Omega^{\epsilon_{(ikl_n)l_{n-1}}}_{(i^+ikl_n)} \dots
\Omega^{\epsilon_{(ikl_1)l_n}}_{(i^+ikl_1)}.
\end{equation}

\noindent Here $\epsilon_{(ikl)m}$ = $\pm 1$ is some sign function. Thereby we find
contribution of the $t$-like triangles and thus $\pmmsig$ in terms of independent
(i.e. leaf/diagonal) curvature matrices.

Thus, in order to represent exponential in (\ref{VEV2}) in the form (\ref{-mR-piR})
and thus the measure in the form (\ref{dN1}) it is sufficient to divide the whole set
of links in the uppest 3D leaf into triples forming the triangles (that is, triangles
do not have common edges) and take this set of triangles as ${\cal M}$ in
(\ref{-mR-piR}). In fig.\ref{M-simplex} example of periodic such set ${\cal M}$
(shaded triangles) is shown for periodic simplicial 3D leaf.

\newcounter{N}\unitlength 0.03mm
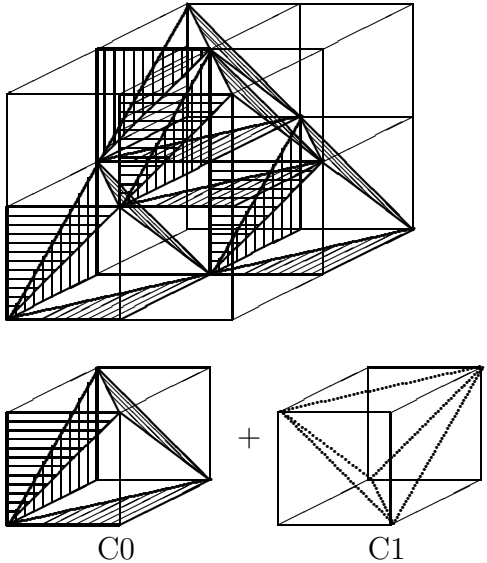
\begin{figure}
\begin{picture}(2000,1600)

\put(0,1000){\line(0,-1){500}}%

\put(0,1000){\line(2,1){800}}%
\put(0,500){\line(2,1){400}}%

\put(800,900){\line(0,-1){500}}%
\put(1000,1000){\line(0,-1){1000}}%

\put(500,500){\line(0,-1){500}}%

\put(0,1000){\line(1,0){500}}%

\put(500,0){\line(1,0){500}}%
\put(800,400){\line(1,0){1000}}%
\put(800,1400){\line(1,0){1000}}%
\put(800,900){\line(1,0){1000}}%

\put(900,1200){\line(1,0){500}}%
\put(500,1000){\line(2,1){800}}%
\put(1000,1000){\line(2,1){800}}%
\put(1400,700){\line(2,1){400}}%

\put(400,200){\line(2,1){400}}%
\put(400,200){\line(1,0){500}}%
\put(1000,0){\line(2,1){400}}%

\put(1300,1400){\line(0,-1){500}}%
\put(1400,1200){\line(0,-1){1000}}%
\put(1800,1400){\line(0,-1){1000}}%

\thicklines

\put(400,700){\line(1,1){500}}%
\put(400,700){\line(1,0){1000}}%
\put(800,1400){\line(1,-1){1000}}%
\put(800,1400){\line(1,-2){200}}%
\put(900,1200){\line(4,-3){400}}%
\put(900,1200){\line(0,-1){1000}}%
\put(400,1200){\line(1,0){500}}%
\put(400,1200){\line(0,-1){1000}}%
\put(400,700){\line(2,1){400}}%
\put(800,1400){\line(0,-1){500}}%
\put(500,1000){\line(0,-1){500}}%
\put(500,1000){\line(1,0){500}}%
\put(0,0){\line(1,1){1000}}%
\put(500,500){\line(2,1){800}}%
\put(0,0){\line(1,0){500}}%
\put(0,0){\line(2,1){400}}%
\put(900,200){\line(1,0){500}}%
\put(500,0){\line(2,1){800}}%
\put(1400,200){\line(2,1){400}}%
\put(0,500){\line(1,0){1000}}%
\put(1000,500){\line(2,1){400}}%
\put(400,700){\line(1,-2){100}}%
\put(1300,900){\line(1,-2){100}}%
\put(500,500){\line(4,-3){400}}%
\put(1000,1000){\line(4,-3){800}}%
\put(0,500){\line(0,-1){500}}%
\put(400,700){\line(1,-1){500}}%
\put(1300,900){\line(0,-1){500}}%
\put(900,200){\line(1,1){500}}%
\put(900,1200){\line(1,-1){500}}%

\thinlines

\put(820,1360){\line(1,-1){400}}%
\put(840,1320){\line(1,-1){300}}%
\put(860,1280){\line(1,-1){200}}%

\multiput(-5,0)(4,7){200}{.}%
\multiput(395,700)(9,2){100}{.}%
\put(-5,1000){.}%
\multiput(495,500)(4,7){100}{.}%

\multiput(495,500)(9,2){100}{.}%
\multiput(-5,0)(9,2){200}{.}%

\multiput(895,200)(4,7){100}{.}%

\setcounter{N}{400}

\multiput(850,700)(-50,0){7}{%
\addtocounter{N}{-40}%
\line(2,1){\value{N}}}

\setcounter{N}{400}

\multiput(800,945)(0,45){7}{%
\addtocounter{N}{-36}%
\line(-2,-1){\value{N}}}

\setcounter{N}{500}

\multiput(500,960)(0,-40){11}{%
\addtocounter{N}{-40}%
\line(1,0){\value{N}}}%

\setcounter{N}{500}

\multiput(860,680)(-40,-20){9}{%
\addtocounter{N}{-50}%
\line(0,1){\value{N}}}%

\setcounter{N}{510}

\multiput(433,1200)(43,0){11}{%
\addtocounter{N}{-43}%
\line(0,-1){\value{N}}}%

\put(920,1160){\line(1,-1){400}}%
\put(940,1120){\line(1,-1){300}}%
\put(960,1080){\line(1,-1){200}}%

\put(1320,860){\line(1,-1){400}}%
\put(1340,820){\line(1,-1){300}}%
\put(1360,780){\line(1,-1){200}}%

\put(420,660){\line(1,-1){400}}%
\put(440,620){\line(1,-1){300}}%
\put(460,580){\line(1,-1){200}}%

\setcounter{N}{400}

\multiput(950,500)(-50,0){7}{%
\addtocounter{N}{-40}%
\line(2,1){\value{N}}}

\setcounter{N}{400}

\multiput(1350,200)(-50,0){7}{%
\addtocounter{N}{-40}%
\line(2,1){\value{N}}}

\setcounter{N}{400}

\multiput(450,0)(-50,0){7}{%
\addtocounter{N}{-40}%
\line(2,1){\value{N}}}

\setcounter{N}{500}

\multiput(0,460)(0,-40){11}{%
\addtocounter{N}{-40}%
\line(1,0){\value{N}}}%

\setcounter{N}{500}

\multiput(900,660)(0,-40){11}{%
\addtocounter{N}{-40}%
\line(1,0){\value{N}}}%

\setcounter{N}{500}

\multiput(1260,380)(-40,-20){9}{%
\addtocounter{N}{-50}%
\line(0,1){\value{N}}}%

\setcounter{N}{500}

\multiput(360,180)(-40,-20){9}{%
\addtocounter{N}{-50}%
\line(0,1){\value{N}}}%

\end{picture}

\begin{picture}(2200,900)

\thicklines

\put(0,0){\line(1,1){500}}%
\put(0,0){\line(2,1){400}}%
\put(0,0){\line(1,0){500}}%
\put(0,0){\line(0,1){500}}%
\put(0,500){\line(1,0){500}}%
\put(500,0){\line(2,1){400}}%
\put(400,200){\line(0,1){500}}%
\put(400,700){\line(1,-2){100}}%
\put(400,700){\line(1,-1){500}}%
\put(500,500){\line(4,-3){400}}%

\thinlines

\put(0,500){\line(2,1){400}}%
\put(500,500){\line(2,1){400}}%
\put(400,200){\line(1,0){500}}%
\put(400,700){\line(1,0){500}}%
\put(500,0){\line(0,1){500}}%
\put(900,200){\line(0,1){500}}%

\put(420,660){\line(1,-1){400}}%
\put(440,620){\line(1,-1){300}}%
\put(460,580){\line(1,-1){200}}%

\setcounter{N}{500}

\multiput(360,180)(-40,-20){9}{%
\addtocounter{N}{-50}%
\line(0,1){\value{N}}}%

\setcounter{N}{500}

\multiput(0,460)(0,-40){11}{%
\addtocounter{N}{-40}%
\line(1,0){\value{N}}}%

\setcounter{N}{400}

\multiput(450,0)(-50,0){7}{%
\addtocounter{N}{-40}%
\line(2,1){\value{N}}}

\multiput(-5,0)(9,2){100}{.}%
\multiput(-5,0)(4,7){100}{.}%

\put(1200,0){\line(2,1){400}}%
\put(1200,500){\line(2,1){400}}%
\put(1700,0){\line(2,1){400}}%
\put(1700,500){\line(2,1){400}}%
\put(1200,0){\line(1,0){500}}%
\put(1200,500){\line(1,0){500}}%
\put(1600,200){\line(1,0){500}}%
\put(1600,700){\line(1,0){500}}%
\put(1200,0){\line(0,1){500}}%
\put(1700,0){\line(0,1){500}}%
\put(1600,200){\line(0,1){500}}%
\put(2100,200){\line(0,1){500}}%

\multiput(1195,500)(18,4){50}{.}%
\multiput(1695,0)(8,14){50}{.}%
\multiput(1695,0)(-7,14){14}{.}%
\multiput(1595,200)(-16,12){25}{.}%
\multiput(1195,500)(11.5,-11.5){44}{.}%
\multiput(1595,200)(11.5,11.5){44}{.}%

\put(1020,350){+}%

\put(400,-150){C0}%
\put(1600,-150){C1}%

\end{picture}

\bigskip
\renewcommand{\baselinestretch}{1.0}
\caption{Example of the set ${\cal M}$ (shaded triangles) with periodic structure such
that any edge does belong to one and only one triangle of ${\cal M}$. The periodic
cell consists of 8 building blocks of the two types C0, C1 alternating in all three
directions. } \label{M-simplex}
\end{figure}
\unitlength 1pt%

Barbero-Immirzi parameter defines area quantization in Loop Quantum Gravity and, in
particular, the number of states on the horizon of black hole. Therefore its possible
value was considered in a number of papers from analysis of the black hole entropy
\cite{rov}-\cite{cor}. Suggested values range from 0.127 \cite{asht} to 0.274
\cite{kh1,cor}. We see that our measure results in asymmetric picture with timelike
areas considerably stronger suppressed than the spacelike ones.

Thus, quantum measure is exponentially suppressed at large areas. The Barbero-Immirzi
parameter proves to play significant role in RC for it provides good behavior of the
measure on areas in the timelike region.

I am grateful to I.B.Khriplovich, who has drawn my attention to importance of the
topological term in the tetrad-connection Lagrangian for gravity, for discussion. The
present work was supported in part by the Russian Foundation for Basic Research
through Grant No. 08-02-00960-a.

\end{document}